\documentclass{article}
\usepackage{spconf,amsmath,graphicx}
\DeclareMathOperator*{\argmin}{arg\,min}
\usepackage{lipsum}
\usepackage{comment}
\usepackage{xcolor}
\usepackage{amssymb}
\usepackage[style=ieee]{biblatex}
\usepackage{tikz}
\usepackage{textcomp}
\usepackage{hyperref}
\newcommand{\detect}[2]{D\left(#1,\,#2\right)}

\newcommand{\norm}[3]{\left|\left| #3 \right|\right|_{#1}^{#2}}
\newcommand{\normsmall}[3]{|| #3 ||_{#1}^{#2}}
\newcommand\copyrighttext{%
  \footnotesize \textcopyright “© 2022 IEEE.  Personal use of this material is permitted.  Permission from IEEE must be obtained for all other uses, in any current or future media, including reprinting/republishing this material for advertising or promotional purposes, creating new collective works, for resale or redistribution to servers or lists, or reuse of any copyrighted component of this work in other works.” }
\newcommand\copyrightnoticeee{%
\begin{tikzpicture}[remember picture,overlay]
\node[anchor=south,yshift=10pt] at (current page.south) {\fbox{\parbox{\dimexpr\textwidth-\fboxsep-\fboxrule\relax}{\copyrighttext}}};
\end{tikzpicture}%
}

\title{Increasing Loudness in Audio Signals: a perceptually motivated approach to preserve audio quality}
\name{A. Jeannerot$^{1,2}$, N. de Koeijer$^{1,3}$, P. Mart\'inez-Nuevo$^{1}$, M. B. M\o ller$^{1}$, J. Dyreby$^{1}$, and P. Prandoni$^{2}$}
  
  \address{$^{1}$ Bang \& Olufsen, Acoustics R\&D, Struer, Denmark \\
      $^{2}$ \'Ecole Polytechnique F\'ed\'erale de Lausanne (EPFL), Lausanne, Switzerland\\
      $^{3}$ Delft University of Technology (TU Delft), Delft, Netherlands}

\begin{document}
%
\maketitle
\copyrightnoticeee

\begin{abstract}
We present a method to maintain the subjective perception of volume of audio signals and, at the same time, reduce their absolute peak value. We focus on achieving this without compromising the perceived audio quality. This is specially useful, for example, to maximize the perceived reproduction level of loudspeakers where simply amplifying the signal amplitude, and hence their peak value, is limited due to already constrained physical designs. In particular, we minimize the absolute peak value subject to a constraint based on auditory masking. This limits the perceptual difference between the original and the modified signals. Moreover, this constraint can be tuned and allows to control the resulting audio quality. We show results comparing loudness and audio quality as a function of peak reduction. These results suggest that our method presents the best trade-off between loudness and audio quality when compared against classical methods based on compression and clipping.
\end{abstract}
\section{Introduction}
\label{sec:introduction}

It is usually desirable for loudspeakers, specially the ones small in size, to be able to increase loudness as much as possible while avoiding any distortion. This may be achieved by simply amplifying the amplitude of the audio signals. However, this is usually limited by physical and electronic design constraints. For example, the maximum displacement of the voice coil of a loudspeaker driver so that it still operates in a linear regime, or the maximum input level of a voltage amplifier so that clipping is avoided. Instead, it is common to resort to processing techniques that lead to an increase of the perceived volume level even in the presence of these limitations. For ease of explanation, we focus on the increase of loudness of voltage signals in this paper. The same principles apply in order to constrain the loudspeaker driver displacement. In this case, a linear model could be readily incorporated \cite{small, theile}.

Indirect methods to increase loudness operate by reducing the amplitude excursions of the audio signal and then, if applicable, reamplifying to increase its energy. There exist two main classical approaches in the literature in this regard: dynamic range compression (DRC) \cite{giannoulis2012digital} and clipping techniques \cite{wathen1958power}. The DRC reduces the amplification gain of the input signal for input levels above a predefined threshold. This reduction is achieved through a compression ratio. Additionally, it incorporates integration periods, referred to as attack and release times, thus introducing memory in the system. Even with a careful choice of these parameters, a DRC may inadvertently lead to poor audio quality \cite{kendrick2015perceived}.

Another common approach is the use of clipping. This is essentially a memoryless input-output map with a linear characteristic around the origin and nonlinear regions for larges input values. Depending on the characteristics of the nonlinear regime, it is possible to further divide clipping into hard and soft clipping. Hard clipping is characterized by clamping the output to a constant value when the input exceeds a predefined threshold. Soft clipping makes this transition smoother. Similar to DRCs, clipping typically results in audio quality degradation \cite{kendrick2015perceived}. In \cite{defraene2012real}, an approach is presented to reduce the impact of clipping by using a simplified masking model. They focus on minimizing the perceptual distance, based on noise masking, between the input and output signals such that the values of the latter lie within a predefined range.

In this paper, we present a perceptually motivated approach that increases the loudness of an audio signal while controlling the degradation in audio quality. As opposed to previous methods, even those incorporating perceptual features, we focus on minimizing the peak value of the input signal under a constraint based on the perceptual detectability of the resulting distortion. This constraint, based on multitonal masking, includes a parameter that is perceptually interpretable, referred to as \textit{detectability}. This provides us with a way of tuning and controlling the output audio quality explicitly. This is particularly relevant in an environment that needs guarantees on the perceived audio quality of the output signals.

The paper is structured as follows: in section \ref{sec:reduction}, we place our proposed perceptually-optimized loudness increase method and classical clipping techniques within an optimization framework. Section \ref{sec:perceptual} introduces the perceptual model used. Finally, in section \ref{sec:simulations}, we demonstrate the performance of the proposed method by using loudness, audio quality, and crest factor reduction as evaluation criteria. We use a dataset of music clips consisting of kick drums and compare against DRCs and clipping techniques.

\section{Loudness Increase: An Optimization Perspective}
\label{sec:reduction}
Instead of directly optimizing for loudness, it is usually more tractable to resort to indirect methods. The perception of volume is not only driven by the peak value of sound pressure waves but, to a large extent, it is also determined by their energy. Thus, in a scenario where maximum amplitude levels are limited, it may be possible to increase loudness by both minimizing the peak value and maximizing the energy. Later, it is always possible to reamplify the signal to its maximum amplitude level, thus increasing loudness. This can be formalized in terms of the so-called crest factor of a signal $\mathbf{x}$, i.e., $\textrm{CR} := 10\log_{10}(||\mathbf{x}||_\infty/||\mathbf{x}||_2)$ where $\mathbf{x}\in\mathbb{R}^N$ for some $N>0$.

Unfortunately, minimizing the crest factor entails solving a nonconvex problem. In order to have stronger theoretical guarantees, we can settle instead for a suboptimal problem that is actually convex. We can achieve this by minimizing the $\ell_\infty$-norm subject to be close to the original signal in an energy sense. Thus, we minimize the peak value and aim at preserving the energy. We show in this section how this approach leads to the classical notions of hard clipping. Finally, our proposed method naturally follows: the energy constraint is weighted by a perceptual criterion instead. This constraint is known as detectability and provides us with control over the resulting audio quality.

\subsection{Non-perceptual Optimization Approach}
Minimizing the crest factor involves minimizing the $\ell_\infty$-norm of a signal and, at the same time, maximizing the $\ell_2$-norm. However, it is more tractable to settle for suboptimal methods. Instead, it is possible to constrain error, in the $\ell_2$ sense, and minimize the peak of a signal. In particular, we can write
\begin{equation}
\label{eq:hard_clipping}
    \begin{array}{rrclcl}
        \displaystyle S_\lambda(\mathbf{x}_0) := \argmin_{\mathbf{x}\in\mathbb{R}^N} & \multicolumn{3}{l}{||\mathbf{x}||_{\infty}+\frac{1}{2\lambda}||\mathbf{x}-\mathbf{x}_0||_{2}}\\
    \end{array}
\end{equation}
for a signal $\mathbf{x}_0\in\mathbb{R}^N$ and some $\lambda>0$. Interestingly, this corresponds to performing hard clipping on $\mathbf{x}_0$. The connection comes from the fact that $S_\lambda(\mathbf{x}_0)$ is the proximal operator of $\lambda||\cdot||_\infty$.

In order to show this, notice that hard clipping can be computed from the soft-thresholding operator $T_\lambda$ \cite{beck2017first}, i.e., $S_\lambda(\mathbf{x}_0) = \mathbf{x_0} - T_\lambda(\mathbf{x}_0)$ where $T_\lambda(x):=\big||x|-\lambda\big|\cdot\textrm{sgn}(x)$ for $|x|>\lambda$ and $T_\lambda(x):=0$ otherwise. Note that $T_\lambda(\mathbf{x})\equiv(T_\lambda(x_i))_{i=1}^{N}$. It follows then that $S_\lambda(x)=x$ for $|x|<\lambda$ and $S_\lambda(x)=\lambda$ otherwise. Similarly, $S_\lambda(\mathbf{x})\equiv(S_\lambda(x_i))_{i=1}^{N}$. 

As shown later, the main issue with this approach is that it leads to poor audio quality. However, we show in the next section how it is possible to include a frequency-domain weighting in the energy difference in such a way that this degradation in audio quality is controlled.

\subsection{Perceptually Motivated Approach}
Instead of simply focusing on solving (\ref{eq:hard_clipping}), we resort to frequency masking. In particular, we introduce the matrix $\mathbf{P}_{\mathbf{x}_0}$ to compute the perceptual difference, or error, between the original and modified signals \cite{van2005perceptual}. This error is constrained by a parameter $c\geq0$, referred to as detectability, that provides information about how noticeable the difference is. This results in the following optimization problem
\begin{equation}
\begin{array}{rrclcl}
\displaystyle \min & \multicolumn{3}{l}{||\mathbf{x}||_{\infty}}\\
\textrm{s.t.} & ||\mathbf{P}_{\mathbf{x}_0}\mathbf{W}(\mathbf{x}-\mathbf{x}_0)||_2 \leq c& & \\
\end{array}
\label{eq:perceptual_opti_1}
\end{equation}
where $\mathbf{W}\in\mathbb{R}^{N\times N}$ is the Discrete Fourier Transform (DFT) matrix. We explain in detail in the next sections how $\mathbf{P}_{\mathbf{x}_0}$ is computed and what the detectability parameter $c$ implies. This form of the optimization problem is particularly interesting since the detectability provides us with a direct way of controlling the audio quality, reducing the peak value, and potentially increasing loudness. We also explore in this paper an alternative formulation, inspired by \cite{defraene2012real}, i.e., 
\begin{equation}
\begin{array}{rrclcl}
\displaystyle \min & \multicolumn{3}{l}{||\mathbf{P}_{\mathbf{x}_0}\mathbf{W}(\mathbf{x}-\mathbf{x}_0)||_2}\\
\textrm{s.t.} & ||\mathbf{x}||_{\infty} \leq \lambda& & \\
\end{array}
\label{eq:perceptual_opti_2}
\end{equation}
where $\lambda>0$ is the maximum amplitude level allowable in the system.

\section{Psychoacoustic Model}
\label{sec:perceptual}
We use a model of human auditory perception that exploits tonal masking. In particular, the distortion detectability $\detect{\mathbf{x}_0}{\mathbf{e}}$ predicts how noticeable a sinusoidal distortion $\mathbf{e}\in\mathbb{R}^N$ is in the presence of a masking signal $\mathbf{x}_0\in\mathbb{R}^N$ \cite{van2005perceptual}. This distortion detectability can be expressed in a compact form as an $\ell_2$-norm in the frequency domain~\cite{taal2012low}:
\begin{equation}
    \detect{\mathbf{x}_0}{\mathbf{e}} := \norm{2}{2}{\mathbf{P}_{\mathbf{x}_0}\hat{\mathbf{e}}}
    \label{eq:detectability}
\end{equation}
where $\hat{\mathbf{e}}:=\mathbf{W}\mathbf{e}$ and $\mathbf{P}_{\mathbf{x}_0}\in\mathbb{R}^{N \times N}$ is a diagonal matrix that weighs the frequency components of $\hat{\mathbf{e}}$ based on the psychoacoustic masking properties of $\hat{\mathbf{x}}_0$. Thus, increasing values of $D$ indicate a more noticeable distortion. This matrix $\mathbf{P}_{\mathbf{x}_0}$ is constructed in the following manner~\cite{taal2012low}:
\begin{equation}
   \mathbf{P}_{\mathbf{x}_0} := \mathrm{diag}\left(\sqrt{ \sum_{i=1}\frac{c_s\hat{\mathbf{h}}^2_i}{\normsmall{2}{2}{\hat{\mathbf{h}}_i\hat{{\mathbf{x}_0}}} + c_a}}\right)
   \label{eq:perceptual_matrix}
\end{equation}
where $\hat{\mathbf{h}}_i\in\mathbb{R}^N$ represents the $i^\mathrm{th}$ filter in a Gammatone filter bank modeling the outer- and middle-ear transfer functions as well as the $i$-th auditory filter (note that $\cdot^2$ operates element-wise). The constants $c_a$ and $c_s$ are calibration coefficients chosen such that $\detect{\mathbf{x}_0}{\mathbf{e}} = 1$, i.e., when $\mathbf{e}$ is just detectable in the presence of $\mathbf{x}_0$~\cite{van2005perceptual}.

In our setting, we constrain the value of $D(\mathbf{x}_0,\mathbf{x}-\mathbf{x}_0)$. In particular, we want to find a signal $\mathbf{x}$ with a lower peak value in magnitude such that the detectability of the error signal $\mathbf{x}-\mathbf{x}_0$ is limited. We interpret the detectability of this error as the perceptual difference between the original and modified signals. As shown in the next section, although the resulting signal may have lower energy as well as maximum amplitude value, its crest factor is reduced and loudness is increased if we add a simple reamplification factor.

\section{Simulations and results}
\label{sec:simulations}
 We focus on enhancing signals mainly composed of low-frequency content. This is of particular interest in reproduction systems since the human hearing system is less sensitive to this frequency range. Our analysis focuses on kick drum sounds sampled from a drum kit, which are characterized by a low-pass spike-like waveform followed by a resonant part. Clearly, naively increasing their loudness by linear amplification can lead to saturation of the voltage amplifier. Since they are often repeated several times across a song, focusing on single kick drum clips is a first step toward increasing the loudness of longer samples. In our experiments, we apply the method proposed in both formulations, as in (\ref{eq:perceptual_opti_1}) and (\ref{eq:perceptual_opti_2}), to 16 clips taken from \cite{kickdataset} with a maximum duration of 1~s and downsampled to 1~kHz. We assess the performance using loudness, audio quality, and crest factor reduction as function of peak reduction as the underlying physical problem is the reduction of the loudspeaker excursion. The dataset used to generate the plots can be accessed online at \cite{alix_jeannerot_2022_5828375}. 

\subsection{Crest Factor Gain}
\begin{figure}[!h]
    \centering
    \includegraphics[width=\linewidth]{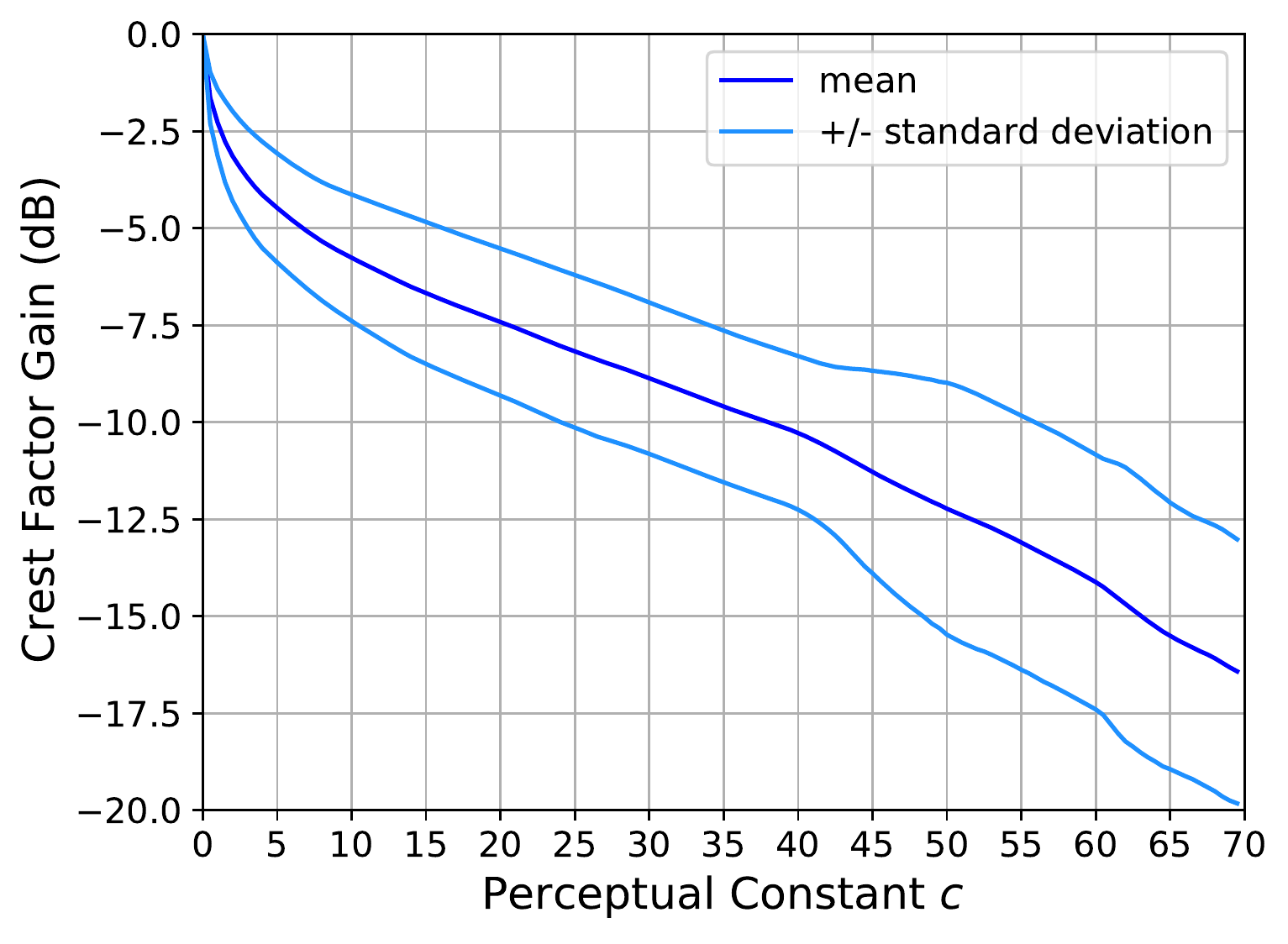}
    \caption{Crest factor reduction for the method proposed in (\ref{eq:perceptual_opti_1}) as a function of the detectability parameter $c$. Results are generated using the 16 clips taken from \cite{kickdataset}.}
    \label{fig:CR_avg_gain}
\end{figure}

From a physical system design perspective, crest factor figures are commonly used to estimate loudspeakers power handling since they provide a measure of absolute peak values relative to RMS \cite{aes_standard}. We introduced the crest factor as a suboptimal metric for increasing loudness. Fig.~\ref{fig:CR_avg_gain} shows the average decrease in crest factor, as computed by (\ref{eq:perceptual_opti_1}), for 16 kick drum clips taken from the dataset. It can be observed that even for small values of detectability, i.e., less than 10, the reduction can lead to a significant increase in perceived volume. 

It is important to emphasize that we introduced a slight change in the matrix $\mathbf{P}_{\mathbf{x}_0}$ in all our simulations. The model from \cite{van2005perceptual} assigns weights close to zero to frequencies below 30~Hz. This can cause the optimization problems in (\ref{eq:perceptual_opti_1}) and (\ref{eq:perceptual_opti_2}) to encourage solutions with energy in an inaudible frequency range. Thus, we assign these weights large values so that they are effectively ignored. 

\begin{figure}[!h]
    \centering
    \includegraphics[width=1.1\linewidth]{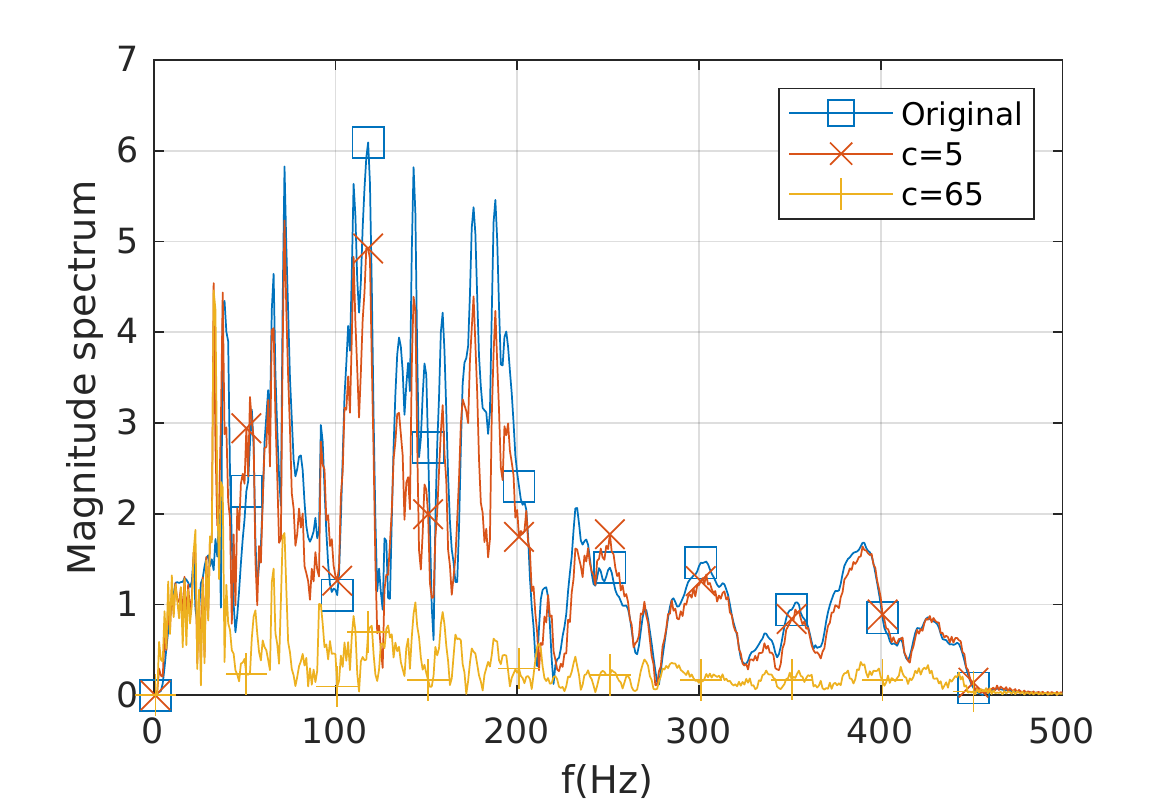}
    \caption{Magnitude spectrum (linear) for three different values of the detectability parameter $c$ in (\ref{eq:perceptual_opti_1}). Results generated for the clip n°1 of \cite{alix_jeannerot_2022_5828375}}
    \label{fig:spectra}
\end{figure}

Fig.~\ref{fig:spectra} shows the magnitude of the spectrum of a single modified kick drum clip from the dataset for different values of the detectability parameter. For low values of this parameter, the constraint on the perceptual difference has more influence. Thus, frequencies between 100 and 200 Hz, which are less constrained by the psychoacoustic model \cite{van2005perceptual}, are modified the most. For larger values of detectability, reducing the absolute peak value dominates and tends to produce a line spectrum. Additionally, the solution tends to present a reduction in energy that can be easily solved by linear amplification.

\subsection{Loudness and Audio Quality}
We compare both formulations of the method proposed, i.e., (\ref{eq:perceptual_opti_1}) and (\ref{eq:perceptual_opti_2}), against a dynamic range compressor and clipping techniques. The dynamic range compressor is implemented with an attack of 0~s, a release of 0.5~s, a ratio of 8, and a knee of 0~dB \cite{giannoulis2012digital}. The hard clipper is easily implemented by clamping to a constant value samples above a threshold. The soft clipper is implemented by setting the appropriate parameters of a dynamic range compressor, i.e., attack and release times of 0~s, a ratio of 50, and a knee of 20dB.

In order to evaluate the performance of these methods, we use loudness and audio quality metrics. We use the Loudness Units relative to Full Scale (LUFS) \cite{ebu}, also known as Loudness, K-weighted, relative to full scale (LKFS)\cite{itu}. This is commonly used, for example, to comply with the audio normalization requirements of many broadcast standards. In terms of audio quality, we use, as an objective measure, the Perceptual Evaluation of Audio Quality (PEAQ) standard \cite{recommendation20011387, thiede2000peaq}. We use 16 kick drum clips from the dataset in \cite{kickdataset}. Note that loudness measurements specially penalize low-frequency content, thus using kick drum clips represents a challenging scenario for all these methods.

We report both loudness and objective audio quality scores as a function of peak reduction. In particular, for each of the methods evaluated, the $x$-coordinate values are the average, across kick drum clips, of the absolute peak value reduction in percent. Similarly, the corresponding $y$-coordinate values are the average, across kick drum clips, of the performance metrics, i.e., loudness or quality scores estimates. As a preprocessing stage for the loudness estimates, the samples were periodically replicated in time and truncated to 1~s. Regarding the PEAQ scores, the reference signal was processed in the same manner as for the loudness estimates. Additionally, the modified audio clips were then appropriate scaled so that they have the same loudness as the reference.

\begin{figure}
    \centering
    \includegraphics[width=\linewidth]{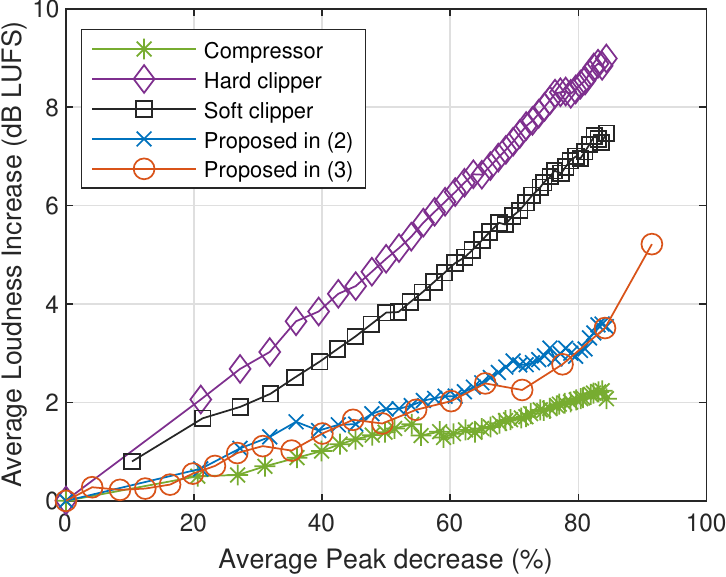}
    \caption{Average loudness increase as a function of average decrease of the absolute peak value.}
    \label{fig:loudness_peak}
\end{figure}

\begin{figure}[!h]
    \centering
    \includegraphics[width=\linewidth]{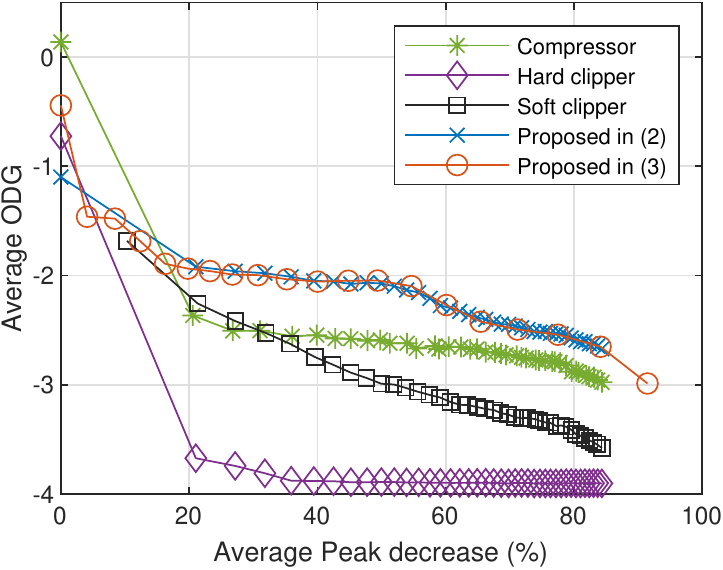}
    \caption{Average Objective Difference Grade (ODG)---i.e., objective sound quality score estimated by PEAQ---as a function of average decrease of the absolute peak value.}
    \label{fig:peaq}
\end{figure}

Fig.~\ref{fig:loudness_peak} shows the loudness values as a function of average peak decrease, i.e., $(||\mathbf{x}_0||_\infty-||\mathbf{x}||_\infty)/||\mathbf{x}_0||_\infty$ in percentage values. We can observe that the hard and soft clippers outperform the rest of the methods in terms of loudness. However, as can be seen in Fig.~\ref{fig:peaq}, they achieve this at the expense of audio quality. Both formulations of the method proposed in this paper have similar performance, i.e., they present the highest objective score in terms of audio quality and, at the same time, they present a significant increase in loudness. These results also suggest that the dynamic range compressor presents a better tradeoff than clipping techniques in terms of loudness and audio quality.

The main difference between both formulations of the method proposed in this paper is that it is possible to either control the resulting audio quality, as in (\ref{eq:perceptual_opti_1}), or the maximum peak value of the resulting signal, as in (\ref{eq:perceptual_opti_2}). The former has the advantage of being able to provide guarantees in terms of resulting audio quality in a system design process.

\section{Conclusion}
We presented a method designed to increase loudness while reducing the absolute peak value of audio signals and, at the same time, preserving audio quality. We achieved this by exploiting a psychoacoustic model, based on multitonal masking, in order to explicitly control either the resulting audio quality or the resulting peak value depending on the formulation of the method. Our results suggest that the methods proposed present a better tradeoff between loudness and audio quality than dynamic range compressors and clipping techniques and should be validated by subjective listening test.

\printbibliography

\end{document}